\documentclass[12pt]{article}
\usepackage[a4paper, margin=2.cm]{geometry}
\usepackage{amsmath}
\usepackage{graphicx}
\usepackage{authblk}
\usepackage{setspace}
\usepackage{verbatim}
\usepackage{color,soul}
\usepackage{hyperref}
\hypersetup{
  colorlinks   = true, 
  urlcolor     = blue, 
  linkcolor    = blue, 
  citecolor    = blue  
}

\DeclareGraphicsExtensions{.png,.pdf}
\title{Reactive molecular dynamics approach to PFAS plasma oxidation in water}

\author[1,2]{Axel Richard}
\author[,1,2]{Pascal Brault \thanks{Corresponding author: pascal.brault@ms4all.eu; pascal.brault@cnrs.fr}}
\author[1]{Nicolas Froloff}
\author[1]{Olivier Aubry}
\author[1]{Dunpin Hong}
\author[1]{Herv\'e Rabat}
\affil[1]{GREMI, Université d'Orléans, CNRS; 14 rue d'Issoudun, F-45067 Orléans, France }
\affil[2]{MS4ALL; 1 avenue du champ de Mars, F-45100 Orléans, France}
\date{\today}

\begin{document}

\maketitle

\section*{Abstract:}
This work establishes a protocol to study via Molecular Dynamics simulation the degradation of Per- and Polyfluoroalkyl Substances (PFAS) in water by hydroxyl radical. To achieve this, molecular dynamics simulations are carried out, using ReaxFF reactive interaction potential. Simulations are carried out under a temperature ramp for determining all possible products. Using this methodology, reaction pathways of perfluorooctanoic acid (PFOA) and perfluorooctanesulfonic acid (PFOS) are identified.\\

\textbf{Keywords:} Molecular Dynamics, simulation, reactive potential, ReaxFF, PFAS, non-thermal-plasma, atmospheric plasma

\textbf{Submitted to:} Open Plasma Science, as a Letter
\newpage

\section{Introduction}

PFAS are fluorinated organic molecules characterized by their carbon chain saturated with fluorine instead of the usual hydrogen atoms. These molecules are widely used in industry, cosmetics, and have numerous applications in everyday products such as non-stick coatings for kitchen utensils \cite{brunn2023}. However, numerous studies on their health impact classify them as potentially carcinogenic \cite{ayodele2024} and reprotoxic \cite{cohen2023}.

Moreover, the carbon-fluorine bond in these molecules is very strong ($>$ 5 eV), leading to a very low degradation rate, earning them the nickname "forever pollutants." Combined with their toxicity, this longevity places them at the center of numerous environmental regulations, particularly on drinking water with the European directive 2020/2184 \cite{eudir2020-2184}, which already lists 20 PFAS to monitor and treat if their concentration exceeds certain thresholds.

Numerous experimental methods are attempted to study PFAS degradation \cite{amen2023} such as non-thermal plasma at atmospheric pressure \cite{singh2019}, ozonation \cite{leung2022}. The common feature of  advanced oxidation processes, including non-thermal plasma at atmospheric pressure, is the availability of the hydroxyl radical HO$^{\bullet}$ in water, which is recognized as a primary oxidant in \cite{lee2010,wang2012}. Despite availability of these experiments, there is a need for predicting the degradation products, pathways and rates of the very large number of PFAS, estimated above 14000 \cite{ackerman2024}. 

Since the basic mechanism of pollutant molecule degradation is a reactive process at molecular scale, molecular dynamics simulation is relevant for predicting degradation products. Molecular dynamics simulations have benefited from the advent of a now popular reactive forcefield: reaxFF \cite{duin2001,senftle2016}. For describing PFAS, a relevant reaxFF forcefield, able to correctly describe carbon - carbon and carbon - fluorine bonds \cite{singh2013}, is available. This potential has also been recently validated on a PFAS sonolysis study \cite{souza2023}. 

\begin{figure}[h!]
    \centering
    \includegraphics[scale=0.6]{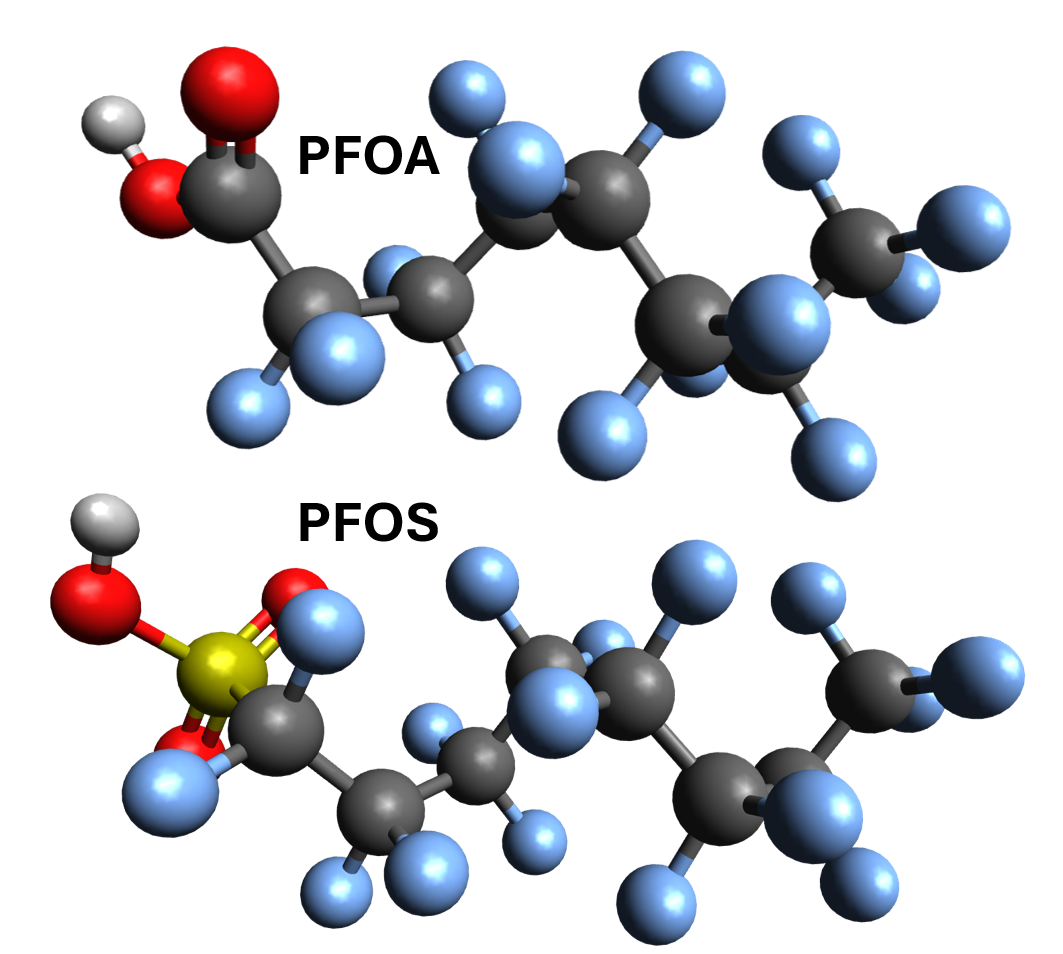}
    \caption{Drawing of PFOA and PFOS. \hl{Color coding: Carbon grey, Hydrogen white, Oxygen red, Fluorine blue, Sulfur yellow}}
    \label{fig:pfoas}
\end{figure}

In the present study, MD simulations were carried out with a temperature ramping between 300 and 5300 K. It allows to  determine the temperature domains for C–C and C–F bond breaking, as well as identifying all possible products up to CO$_2$ conversion under action of HO$^{\bullet}$. \hl{It should be noticed that this work is not corresponding to a specific process, but it is proposing a methodological tool for finding energetic requirements for PFAS degradation.} It is applied to the most known PFAS, say perfluorooctanoic acid (PFOA, C$_8$HF$_{15}$O$_2$) and Perfluorosulfonic acid (PFOS, C$_8$HF$_{17}$O$_3$S). Figure \ref{fig:pfoas} displays the form of these two molecules. 
The highest temperature of the ramp is chosen to correspond to temperature reachable in sonolysis experiments where cavitation bubbles are able to reach such high temperature locally \cite{souza2023}.

\section{Molecular Dynamics simulations}
Molecular Dynamics is a simulation method that describes the behavior of an N-body system. It considers each of the $N$ species in the simulation as distinct objects subject to Newtonian mechanics, i.e., with a defined position and velocity dependent only on the forces exerted on them, according to the following equation:

\begin{equation}
    \label{eq1}
     m_i \frac{d^{2} \vec{r}_i(t)}{d t^{2}} = \vec{f}_i(t), \quad \mbox{with} \quad
     \vec{f}_i(t) =  -\frac{\partial V(\vec{r}_1(t), \vec{r}_2(t), ...,\vec{r}_n(t))}{\partial \vec{r}_i(t)} 
\end{equation}

Where $\vec{r}_{i}(t)$ is the position of atom $i$ at time $t$, $m_{i}$ is its mass, and $\vec{f}_{i}(t)$ are the forces applied to atom $i$ at time $t$. $V = V(\vec{r}_1, \vec{r}_2, ...,\vec{r}_n)$ is the interaction potential among all species.

These forces exerted on each atom then determine its position and velocity at the next time step through discrete time integration. By repeating this method as many times as necessary, it becomes possible to determine the behavior of atom  assembly.

To correctly simulate the behavior of these systems, we need to know the initial configuration, preferably matching, at best, experimental situations. The forces between atoms are derived from the interaction potential $V(\vec{r}_1(t), \vec{r}_2(t), ...,\vec{r}_n(t))$. There are various categories of potentials. \emph{Ab-initio} potentials are based on quantum mechanics equations which require significant computational power. There is also a category of semi-empirical potentials that approximate these forces, allowing much shorter computation times for the same simulation. One of the most relevant is the reaxFF one, which is both reactive (allows bond breaking and formation based on bond order concept) and includes variable charges (atomic fractional charges calculated and equilibrated every chosen timestep interval) \cite{senftle2016}.  

 ReaxFF was  originally developed to simulate hydrocarbons \cite{duin2001} and has undergone numerous evolutions applying it to a broader range of simulations, including oxidation in aqueous environments \cite{Yusupov2014,brault2021}. This potential considers numerous factors when calculating the energy contribution of the interaction between each atom in the system. The system's energy is calculated as follows:
\[
E_{system} = E_{bond} + E_{over} + E_{under} + E_{val} + E_{pen} + E_{tors} + E_{conj} + E_{vdW} + E_{Coulomb}
\]
$E_{bond}$ uses the distance between two atoms to determine the bond order and calculate its energy. The terms $E_{over}$ and $E_{under}$ impose energy penalties when an atom has too many or too few bonds with its neighbors. Additionally, $E_{val}$ and $E_{tors}$ add an energy penalty to the system when the valence and torsion angles deviate from the expected equilibrium value, respectively. $E_{pen}$ is present to penalize certain configurations when two double bonds share the same atom. The term $E_{conj}$ is used to account for conjugated systems, while $E_{vdW}$ represents the van der Waals interaction between different atoms. Finally, $E_{Coulomb}$ represents the Coulomb interactions between atoms due to the partial charges assigned to them by the "Electronegativity Equalization Method" \cite{mortier1986,janssens1997}.

For completeness, we used the reaxFF parameters used in  \cite{singh2013,souza2023} which were already tested for describing C-F bonds in various situations, including PFAS.

To study the degradation products of PFAS, a methodology similar to Brault et al. \cite{brault2021} was adopted. The simulation boxes are cubes with sides of 15.3 \r{A} containing, randomly located, a PFAS molecule, 10 HO$^{\bullet}$ radicals, and 30 water molecules.  The 10 HO$^{\bullet}$ radicals allow a concentration relative to the pollutant similar to that found in water when subjected to a plasma discharge on the surface. While 100 water molecules would be needed to represent the correct water density, only 30 were used to make the calculations faster, without  influencing the results \cite{brault2023}. \hl{This a very high ratio of PFAS and HO$^{\bullet}$ with water molecules: water molecules are introduced for mimicking the environment of the degradation reactions. This allows accelerating the dynamics for determining the created products. Periodic boundary conditions are applied in all directions, mimicking bulk water processes.}

\begin{figure}[h!]
    \centering
    \includegraphics[scale=0.8]{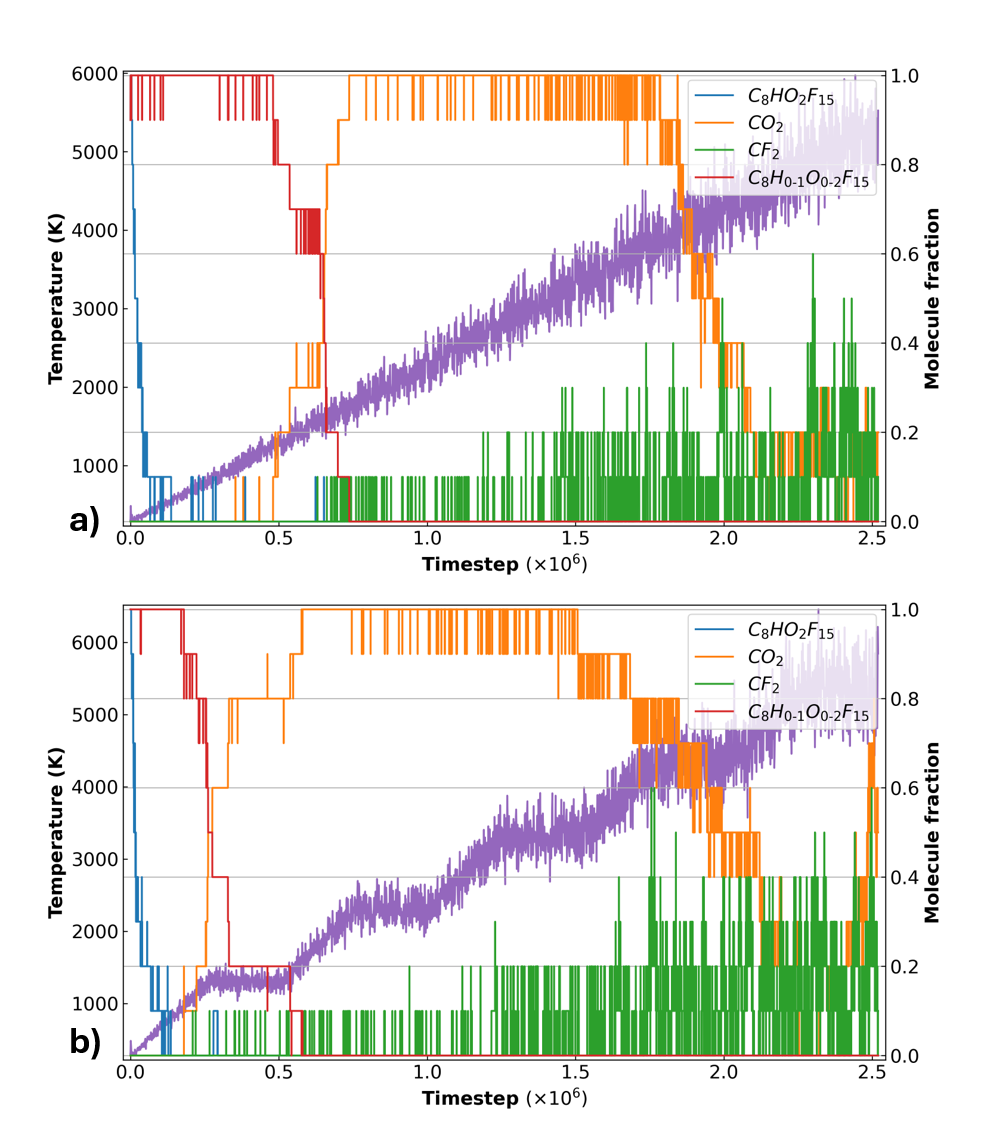}
    \caption{Comparison of a simulation a) without and b) with stepped temperature ramp. In purple is the temperature. The other colors represent the average number of molecules of each species during the simulation. In blue is the original PFOA molecule, in yellow $\mathrm{CO}_{2}$, in green $\mathrm{CF}_{2}$, and in red the sum of molecules with 8 carbons and 15 fluorine atoms. We can see that the two figures are quite similar, with reactions occurring faster in terms of time for the stepped ramp but at similar temperatures.}
    \label{fig:temperature_comparison}
\end{figure}

For each considered PFAS, a set of 10 simulations was performed, with different initial conditions  and then subjected to equilibration at 300 K for 1000 fs with time steps of 0.1 fs, followed by a temperature ramp from 300 K to 5300 K with a rate of 50 fs.K$^{-1}$, ending with a second equilibration step at a constant temperature of 5300 K. \hl{The temperature is controlled using a Nose-Hoover thermostat with a 100 fs damping time.} We then monitor the evolution of the chemical species present during the simulation for each simulation boxes and then results are averaged. An alternative way was tested: the use of a stepped temperature ramp to give more chance to potential species that might not have time to form in the previous case. However, as seen in Figure \ref{fig:temperature_comparison}, there is no noticeable differences. The continuous ramp method was thus retained to easily link integration time steps to temperature.

\section{Results and discussion}
\begin{figure}[h!]
    \centering
\includegraphics[scale=0.5]{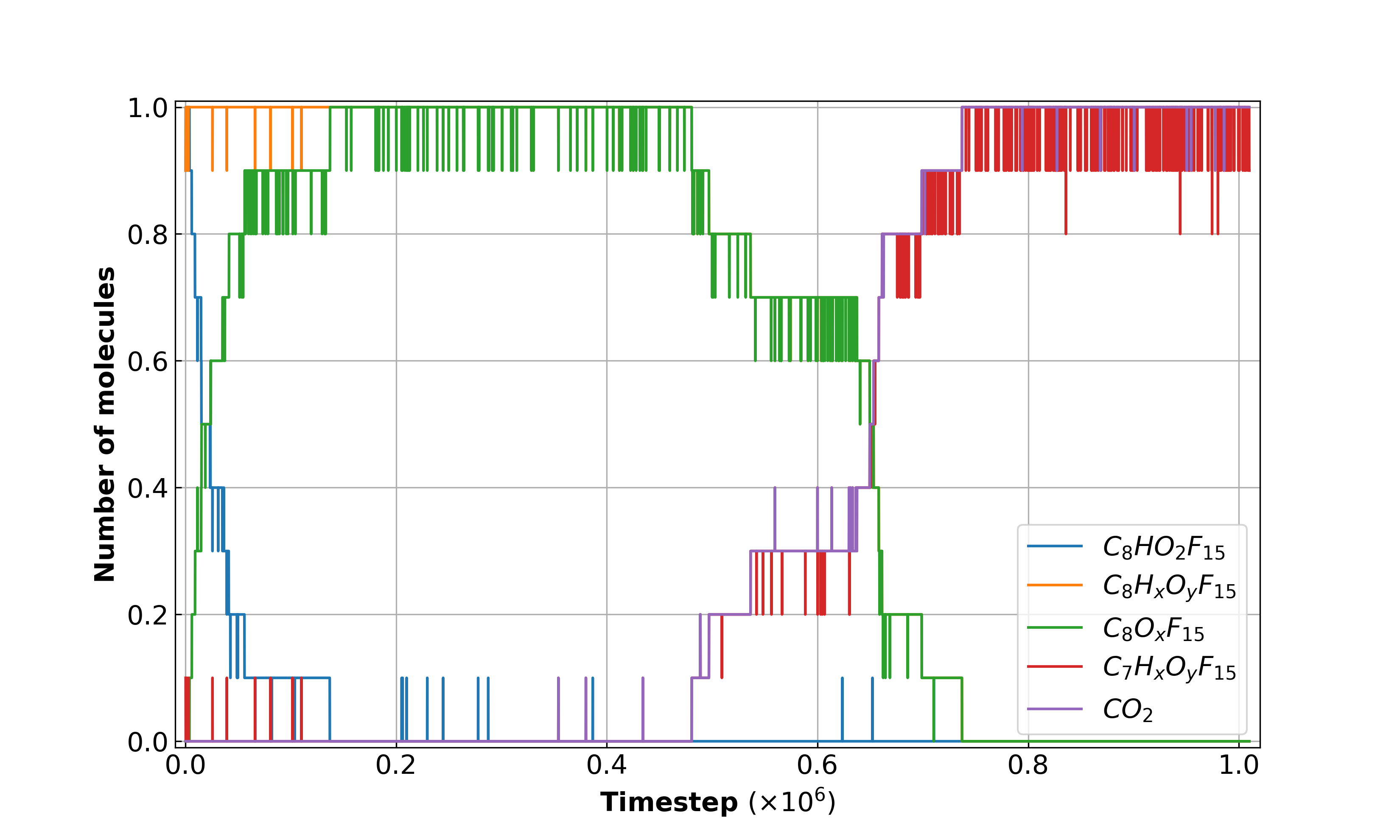}
    \caption{Evolution of the average number of molecules of certain species at the beginning of the simulation. \hl{Table S2 in supplementary information gives the details of the considered molecules}}
    \label{fig:pfoa_start}
\end{figure}

We are interested in the evolution of species during the simulation to determine the reactions that occurred for a PFOA molecule. We average the results over the ten simulations performed to better represent the evolution of various species. Figure \ref{fig:pfoa_start} shows the reaction path taken at the beginning of the simulation. The blue line represents PFOA; we see that this molecule rapidly disappears in favor of  $\mathrm{C}_{8} \mathrm{F}_{15} \mathrm{O}_{2}$ (in green), indicating that the hydrogen atom of the functional group of the PFAS molecule is rapidly removed. The next step is much slower and occurs between 900 K ($\approx$ 3. 10$^5$ timesteps) and 1700 K ($\approx$ 6. 10$^5$ timesteps): $\mathrm{C}_{8} \mathrm{F}_{15} \mathrm{O}_{2}$  molecule disappears while a $\mathrm{CO}_{2}$ molecule (in purple) and a $\mathrm{C}_{7} \mathrm{F}_{15}$ molecule (in red) appear, suggesting that the previous molecule splits. Additionally, the yellow curve closely following the green curve shows that no other molecule containing 8 carbon and 15 fluorine atoms is formed.

\begin{figure}[h!]
    \centering
    \includegraphics[scale=0.5]{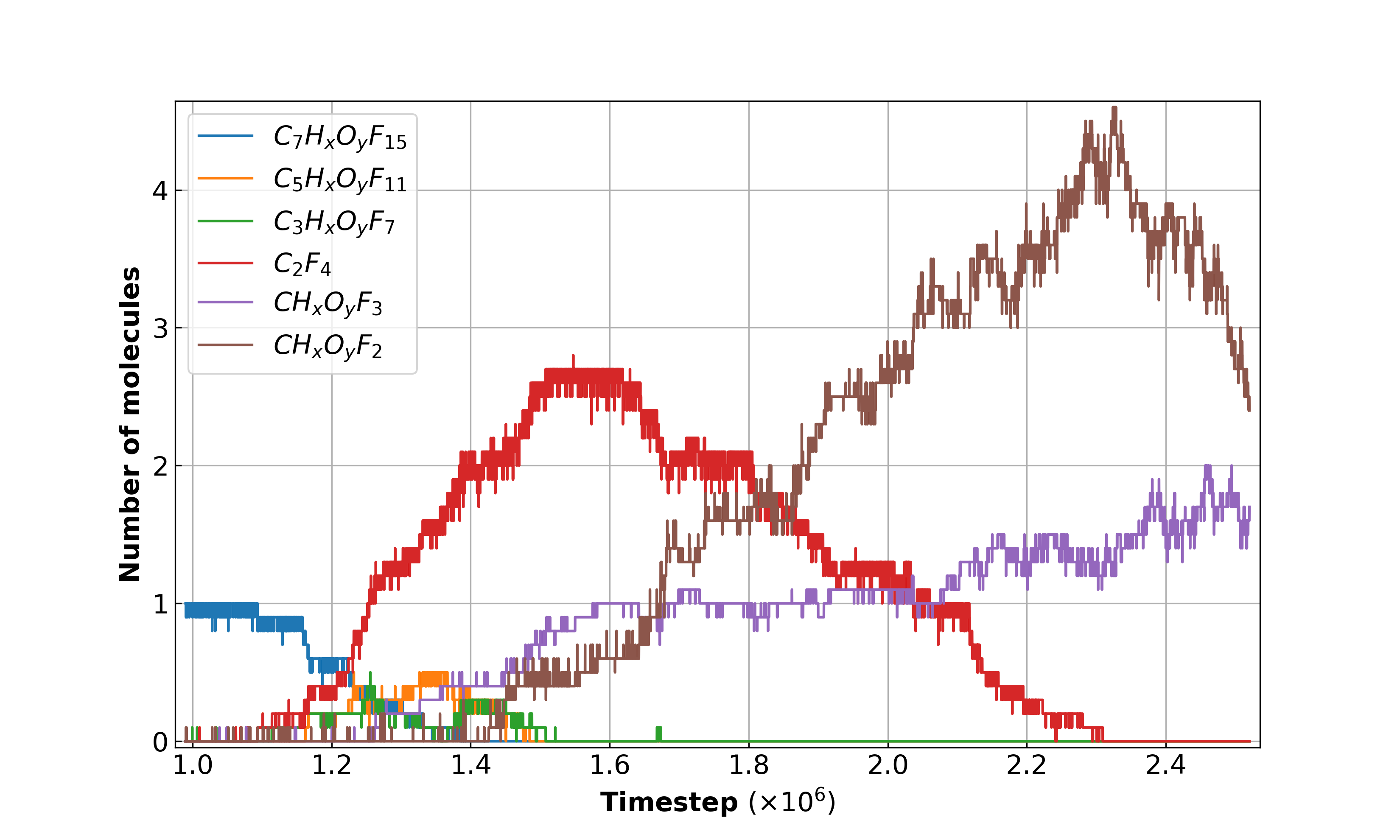}
    \caption{Evolution of the average number of molecules of certain species in the middle of the simulation. \hl{Table S3 in supplementary information gives the details of the considered molecules}}
    \label{fig:pfoa_final}
\end{figure}

In the long term, our main molecule continues to degrade. We can see in Figure \ref{fig:pfoa_final} the continuation of the reaction. Thus, the $\mathrm{C}_{7} \mathrm{F}_{15}$ molecules (still in red) are degraded into  $\mathrm{C}_{5} \mathrm{F}_{11}$ molecules (blue) and  $\mathrm{C}_{2} \mathrm{F}_{4}$ molecules (green), in the temperature range from 2500-3000 K. The $\mathrm{C}_{5} \mathrm{F}_{11}$ molecules are then rapidly degraded into  $\mathrm{C}_{3} \mathrm{F}_{7}$ molecules (yellow) and also to $\mathrm{C}_{2} \mathrm{F}_{4}$ molecules. Then, the $\mathrm{C}_{3} \mathrm{F}_{7}$ molecules are degraded into $\mathrm{C}_{2} \mathrm{F}_{4}$ and $\mathrm{CF}_{3}$ (purple). Finally, the $\mathrm{C}_{2} \mathrm{F}_{4}$ molecules begin to degrade into two $\mathrm{CF}_{2}$ molecules around 3400 K. \hl{All generated molecules during the 10 runs are listed in Table S1 in the supplementary information.}

\begin{figure}[h!]
    \centering
    \includegraphics[scale=0.5]{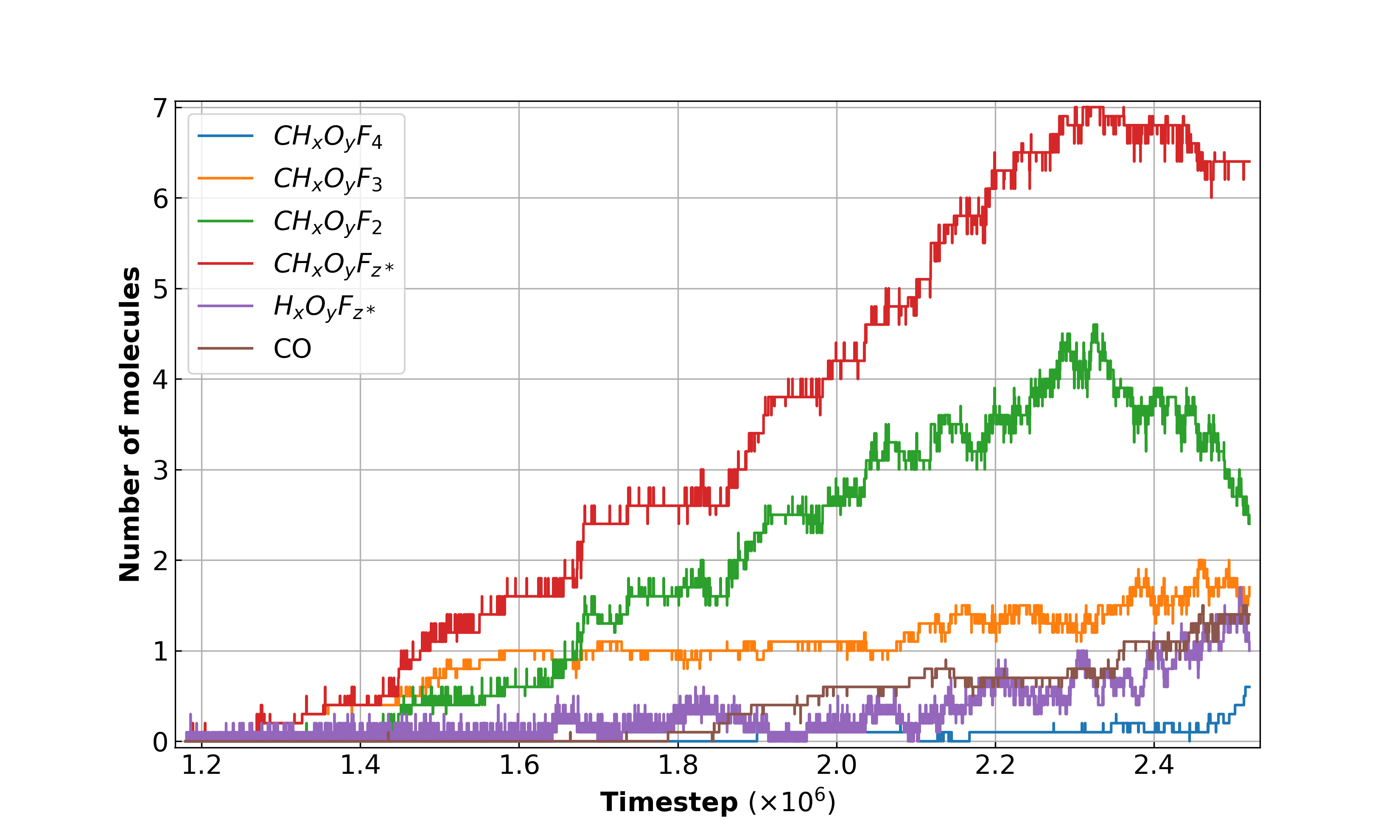}
    \caption{Evolution of the average number of molecules of C1 species at the end of the simulation. x, y = 0, 1, ... and * means z $\ge$ 1. \hl{Table S4 in supplementary information gives the details of the considered molecules}}
    \label{fig:pfoa_endproducts}
\end{figure}
To go further, we identify the products at the end of the reaction in Figure \ref{fig:pfoa_endproducts}. We thus follow the evolution of the number of molecules containing one carbon, an arbitrary number of oxygen and hydrogen, with 4, 3, and 2 fluorine atoms in red, green, and yellow, respectively. The number of $\mathrm{CH}_{x} \mathrm{F}_{3} \mathrm{O}_{y}$ molecules is quite stable from 3500 K, while the number of $\mathrm{CH}_{x} \mathrm{F}_{2} \mathrm{O}_{y}$ continues to increase until 4300 K due to the degradation of $\mathrm{C}_{2} \mathrm{H}_{x} \mathrm{F}_{4} \mathrm{O}_{y}$ molecules seen previously. In the end, we have about 0.2 $\mathrm{CH}_{x} \mathrm{F}_{4} \mathrm{O}_{y}$, 1.2 $\mathrm{CH}_{x} \mathrm{F}_{3} \mathrm{O}_{y}$, 4 $\mathrm{CH}_{x} \mathrm{F}_{2} \mathrm{O}_{y}$. We have represented the sum of molecules containing one carbon and at least one fluorine in blue, and we find the expected 6.4 molecules. Additionally, the number of CO molecules being 1.6, we can be certain of having identified all the degradation products containing carbon found at the end of the simulation. It is worth noting that identifying $\mathrm{CF}_{4}$ as a relatively rare degradation product demonstrates the utility of our method for detecting reactions with a relatively low probability, allowing us to provide a more complete picture of the final and intermediate products that appear during the degradation reaction.

We were able to identify the degradation pathway of a PFOA molecule. Indeed, the molecule first loses its hydrogen, followed by a carbon and two oxygen atoms that form a $\mathrm{CO}_{2}$. Then, the remaining organic chain is cut two carbons at a time, forming numerous $\mathrm{C}_{2} \mathrm{F}_{4}$ molecules, which are then mainly degraded into $\mathrm{CF}_{2}$. We can represent the reaction sequence as follows:

\[
\begin{aligned}
\mathrm{C}_{8} \mathrm{HF}_{15} \mathrm{O}_{2}+\mathrm{HO^{\bullet}} & \longrightarrow \mathrm{C}_{8} \mathrm{F}_{15} \mathrm{O}_{2}+\mathrm{H}_{2} \mathrm{O} \\
\mathrm{C}_{8} \mathrm{F}_{15} \mathrm{O}_{2} & \overset{nHO^{\bullet}}{\longrightarrow} \mathrm{C}_{7} \mathrm{H}_{x}\mathrm{O}_{y} \mathrm{F}_{15}+\mathrm{CO}_{2}  \\
\mathrm{C}_{7} \mathrm{F}_{15} & \overset{nHO^{\bullet}}{\longrightarrow} \mathrm{C}_{5}\mathrm{H}_{x}\mathrm{O}_{y} \mathrm{F}_{11}+\mathrm{C}_{2} \mathrm{F}_{4} \\
\mathrm{C}_{5} \mathrm{F}_{11} & \overset{nHO^{\bullet}}{\longrightarrow} \mathrm{C}_{3} \mathrm{H}_{x}\mathrm{O}_{y}\mathrm{F}_{7}+\mathrm{C}_{2} \mathrm{F}_{4} \\
\mathrm{C}_{3} \mathrm{F}_{7} & \overset{nHO^{\bullet}}{\longrightarrow} \mathrm{C}\mathrm{H}_{x}\mathrm{O}_{y}\mathrm{F}_{3}+\mathrm{C}_{2} \mathrm{F}_{4} \\
\mathrm{C}_{2} \mathrm{F}_{4} & \overset{nHO^{\bullet}}{\longrightarrow} 2 \mathrm{CF}_{2} \\
& x,y=0,1,...; n =0,1, ...
\end{aligned}
\]

This reaction pathways are consistent with literature \cite{leung2022}. Nevertheless, it should be noticed that the formation of C$_{7}$F$_{15}$ can be alternatively followed by the formation of C$_{6}$F$_{15}$COOH. By this way the carboxylic group always remains up to the triflic acid final product CF$_{3}$COOH after interaction with two additional HO$^{\bullet}$ radicals,  H$_2$O and HF \cite{singh2019}. It thus requires a higher concentration of HO$^{\bullet}$ radicals close to the PFOA and related products.

\subsection{Case of PFOS}
The same methodology can be used for perfluorooctanesulfonic acid (PFOS) with the chemical formula $\mathrm{C}_{8} \mathrm{F}_{17} \mathrm{SO}_{3} \mathrm{H}$. We find that the molecule also very quickly loses its hydrogen with a HO molecule to form a water molecule. Then, the molecule separates into two new molecules at a temperature ranging from 1700 K to 2300 K. $\mathrm{SO}_{3}$ and $\mathrm{C}_{8} \mathrm{F}_{17}$ molecules are thus formed. The $\mathrm{C}_{8} \mathrm{F}_{17}$ molecule will then yield twice a molecule of $\mathrm{C}_{2} \mathrm{H}_{4}$ when the temperature ranges from 2500 K to 3300 K. The resulting $\mathrm{C}_{4} \mathrm{F}_{9}$ molecule is then degraded into three different molecules: $\mathrm{C}_{2} \mathrm{F}_{4}$, $\mathrm{CF}_{2}$, and $\mathrm{CF}_{3}$ from 3100 K. At the end of the simulation, $\mathrm{SO}_{3}$ (sometimes accompanied by one or two water molecules) and numerous CF and $\mathrm{CF}_{2}$ are obtained. \hl{All generated molecules during the 10 runs are listed as Table S5 in the suplementary information.}

\begin{figure}[h!]
    \centering
\includegraphics[scale=0.5]{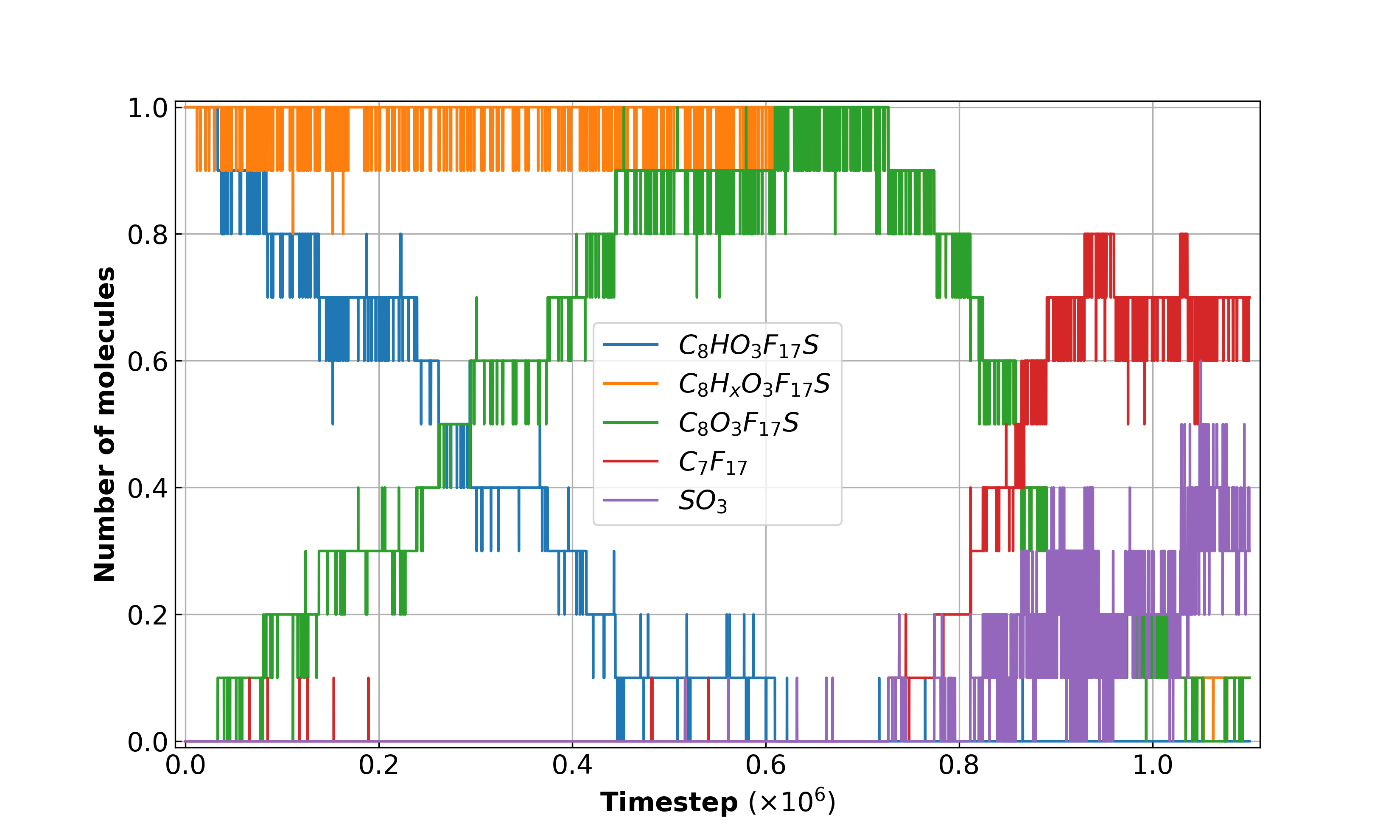}
    \caption{Evolution of the average number of PFOS and product molecules at the beginning of the simulation, up to 0.1 ns. \hl{Table S6 in supplementary information gives the details of the considered molecules}}
    \label{fig:pfos_start}
\end{figure}
\begin{figure}[h!]
    \centering
    \includegraphics[scale=0.5]{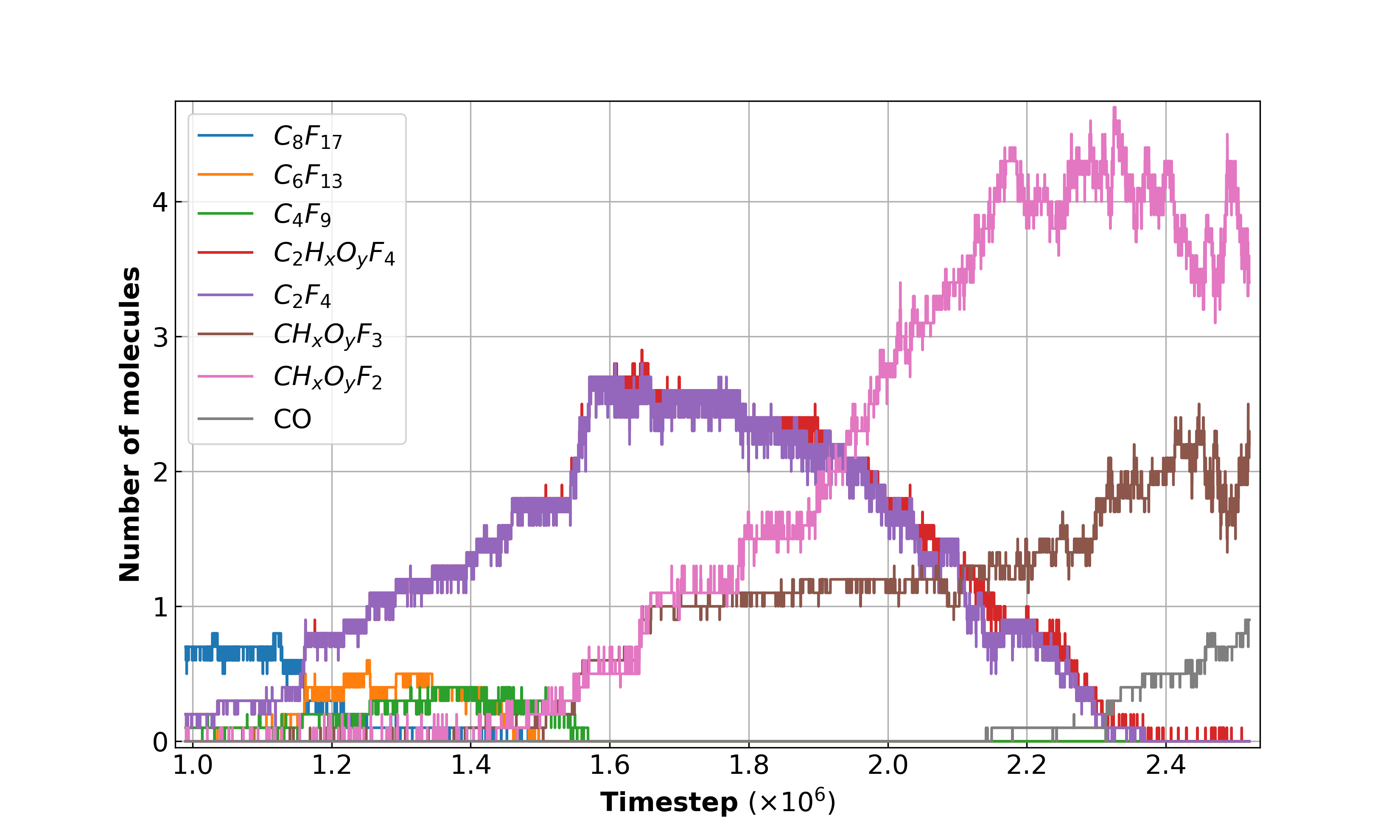}
    \caption{Evolution of the average number of product molecules  between 0.1 and 0.25 ns of the simulation. \hl{Table S7 in supplementary information gives the details of the considered molecules}}
    \label{fig:pfos_final}
\end{figure}
\begin{figure}[h!]
    \centering
    \includegraphics[scale=0.5]{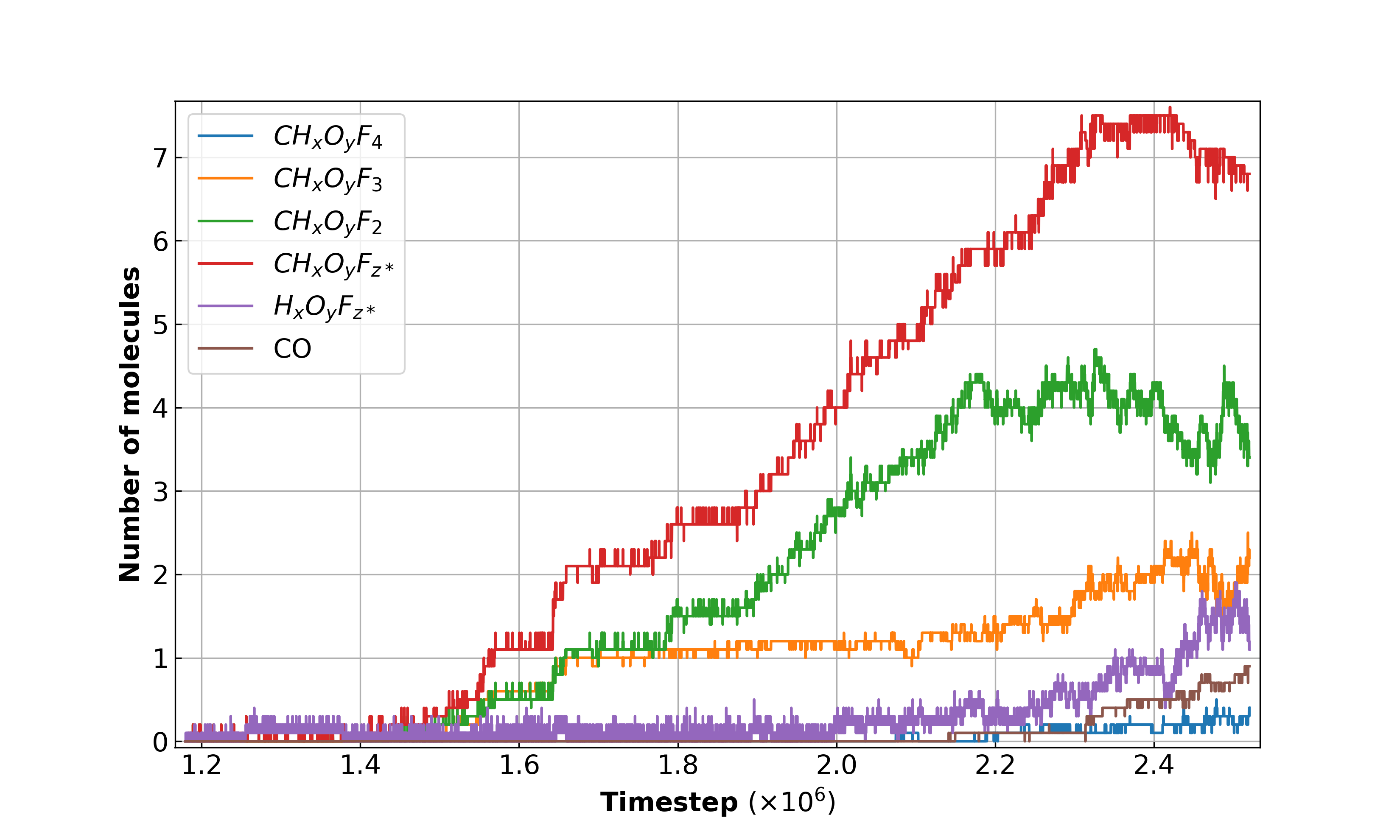}
    \caption{Evolution of the average number of molecules of C1 species at the end of the simulation. x, y = 0, 1, ... and * means z $\ge$ 1. \hl{Table S8 in supplementary information gives the details of the considered molecules}}
    \label{fig:pfos_endproducts}
\end{figure}
Thus, the reaction pathways extracted from figures \ref{fig:pfos_start},\ref{fig:pfos_final} and \ref{fig:pfos_endproducts} consistent with previous experimental findings \cite{leung2022}, can be written as follows:

\[
\begin{aligned}
\mathrm{C}_{8} \mathrm{HF}_{17} \mathrm{O}_{3} \mathrm{S}+\mathrm{HO^{\bullet}} & \longrightarrow \mathrm{C}_{8} \mathrm{F}_{17} \mathrm{O}_{3} \mathrm{S}+\mathrm{H}_{2} \mathrm{O} \\
\mathrm{C}_{8} \mathrm{F}_{17} \mathrm{O}_{3} \mathrm{S} & \overset{nHO^{\bullet}}{\longrightarrow} \mathrm{C}_{8}\mathrm{H}_{x}\mathrm{O}_{y} \mathrm{F}_{17}+\mathrm{S}\mathrm{O}_{3}  \\
\mathrm{C}_{8} \mathrm{F}_{17} &\overset{nHO^{\bullet}}{\longrightarrow} \mathrm{C}_{6} \mathrm{F}_{13}+\mathrm{C}_{2} \mathrm{F}_{4} \\
\mathrm{C}_{6} \mathrm{F}_{13} & \overset{nHO^{\bullet}}{\longrightarrow}\mathrm{C}_{3} \mathrm{H}_{x}\mathrm{O}_{y}\mathrm{F}_{7}+\mathrm{C}_{2} \mathrm{F}_{4} \\
\mathrm{C}_{4} \mathrm{F}_{9} & \overset{nHO^{\bullet}}{\longrightarrow} \mathrm{C}\mathrm{H}_{x}\mathrm{O}_{y}\mathrm{F}+\mathrm{C}\mathrm{H}_{x}\mathrm{O}_{y}\mathrm{F}_{2}+\mathrm{C}_{2}\mathrm{H}_{x}\mathrm{O}_{y} \mathrm{F}_{4} \\
& x,y=0,1,...; n =0,1, ...
\end{aligned}
\]

Since these simulations have been carried out only with HO$^{\bullet}$, it should be noticed that under other physicochemical degradation conditions, including other reactive radicals, accessible by reactive molecular dynamics simulation, the degradation products, reaction paths, rates of appearance and proportions of metabolites produced over time can be very different, for the same starting PFAS. More generally, this makes it possible to consider describing, by this type of simulation, any advanced oxidation/reduction process, including photo-assisted.

\section{Conclusion}
This work allowed us to use a relevant ReaxFF potential for simulating the degradation of PFAS in water by HO radical oxidation, which is expected to mimick the effects of plasma interaction with PFAS polluted water.   It also validates the simulation methodology involving a temperature ramp during the simulation. This methology allows thus to predict all possible degradation products of PFAS interacting with HO$^{\bullet}$ in water. Moreover the link between products and temperature can be useful for reverse analysis: knowing the temperature at which a product is formed can be used to infer the best process to obtain it.

\section*{Acknowledgements}
Conseil R\'egional Centre Val de Loire is gratefully ackowledged for project Perturb'Eau under grant \#2021-00144786.  

\bibliographystyle{unsrt}
\bibliography{ops15754_refs.bib}
\end{document}